\documentclass[%
 reprint,
 amsmath,amssymb,
 aps,
10.5pt]{revtex4-2}
\usepackage{mathrsfs}
\usepackage{siunitx}
\usepackage[dvips]{epsfig}
\usepackage{amsfonts}
\usepackage{color}
\def\bm#1{\mbox{\boldmath{$#1$}}}
\def\rr#1{(\ref{#1})}

\def\cprime{$'$}
\newcommand{\be}{\begin{equation}}
\newcommand{\ee}{\end{equation}}
\usepackage{graphicx}
\usepackage{dcolumn}
\usepackage{bm}

\begin{document}

\preprint{APS/123-QED}

\title{Thermocapillary migrating odd viscous droplets}
\author{A.Aggarwal}
\affiliation{Department of Materials Science \& Engineering, Robert R. McCormick School of Engineering and Applied Science, Northwestern University, Evanston IL 60208 USA and \\Center for Computation and Theory of Soft Materials, Northwestern University, Evanston IL 60208 USA
}
\author{E.Kirkinis}
\affiliation{Department of Materials Science \& Engineering, Robert R. McCormick School of Engineering and Applied Science, Northwestern University, Evanston IL 60208 USA and \\Center for Computation and Theory of Soft Materials, Northwestern University, Evanston IL 60208 USA
}
%
%
\author{M.Olvera de la Cruz}
\email{m-olvera@northwestern.edu}
\affiliation{Department of Materials Science \& Engineering, Robert R. McCormick School of Engineering and Applied Science, Northwestern University, Evanston IL 60208 USA and \\Center for Computation and Theory of Soft Materials, Northwestern University, Evanston IL 60208 USA
}

\date{\today}

\begin{abstract}
A droplet of a classical liquid surrounded by a cold gas placed on a hot substrate is accompanied by unremitting internal circulations, while the droplet remains immobile. Two identical cells with opposite sense of circulation form in the interior due to the thermocapillary effect induced by the gas and substrate temperature difference. Under the same conditions, a droplet composed of an odd viscous liquid exerts a compressive stress on the cell rotating in one sense and tensile on the cell rotating in the opposite sense resulting in a tilted droplet configuration. A sufficiently strong thermal gradient leads the contact angles to overcome hysteresis effects and induces droplet migration.
\end{abstract}

\maketitle

\begin{figure*}
\vspace{-15pt}
\begin{center}
\includegraphics[height=2in,width=7in]{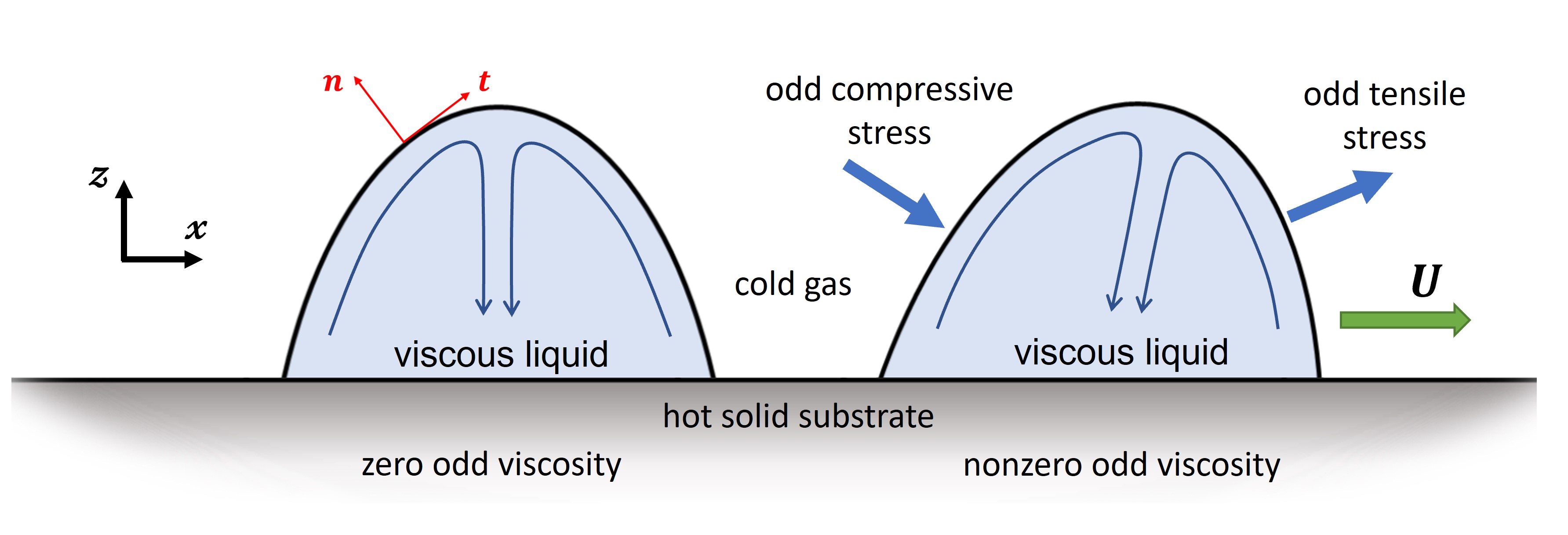}
\end{center}
\vspace{-5pt}
\caption{\textbf{Left:} Two recirculating cells formed by thermocapillarity in a viscous droplet sitting on a 
hot substrate and surrounded by a cold ambient gas phase, are symmetric with respect 
to the center axis of the droplet.
\textbf{Right:} 
Odd viscosity breaks these symmetries and exerts a compressive stress on one cell and a tensile stress on the other
leading to a deformation of the equilibrium shape of the droplet and the onset of migration with velocity $U$. 
By $\mathbf{n}$ and $\mathbf{t}$ we denote the unit normal and tangent 
vectors at the liquid-gas interface. {This same two-dimensional geometry (slice of a 3D droplet) has 
been adopted by many experiments 
and theoretical works on droplet motion.
See Fig. \ref{odd_viscosity_droplet3} for corresponding
results obtained with a \emph{three-dimensional} thermocapillary droplet}. 
\label{fig1} }
\vspace{-5pt}
\end{figure*}


Fluidic devices suffer from a lack of direct sample accessibility and increased friction. On the other hand
free-surface flows, enjoying a large surface-to-volume ratio, can circumvent these 
challenges and lead to effective liquid 
driving through interfacial processes, for example via thermal gradients \cite{Darhuber2005, *Ehrhard1991, *Ehrhard1993, Brochard1989, *Brzoska1993}.

A \emph{nonisothermal}
droplet sitting on solid heated substrated surrounded by a cold ambient gas phase
is accompanied by internal liquid circulations in the form of two identical cells rotating in an opposite
sense (cf. left of Fig.\ref{fig1}).
This is because the thermocapillary effect transports interfacial liquid from
the hot contact lines, where surface tension is lowest, to the cool droplet peak where surface tension is highest. 
Because the liquid
is viscous, it drags the bulk liquid as well. 

Avron \cite{Avron1998} showed that a classical liquid is endowed with an extra viscosity coefficient, 
the odd or Hall viscosity, in the presence of time-reversal symmetry breaking and that the accompanying stress is non-dissipative. The existence of this odd viscous stress was recently established experimentally \cite{Soni2019}. 

In this Letter we show that, odd viscosity induces migration of the aforementioned internally-circulating \emph{nonisothermal droplets}
on a heated solid substrate surrounded by a colder ambient gas phase. 
Because the two cells formed by thermocapillarity have an opposite
sense of circulation, 
odd viscosity applies a compressive stress on one cell and a tensile stress on the other (cf.~right of Fig.~\ref{fig1})
leading to a deformation of the equilibrium shape of the droplet and the onset of migration. 
We
tacitly assume that odd viscosity has already been established in the liquid without referring
explicitly to its underlying mechanism (cf. \cite{Souslov2019}) and proceed to establish
its effects on thermocapillary droplet migration.

\begin{figure}[b]
\vspace{-5pt}
\begin{center}
\includegraphics[height=2.2in,width=3in]{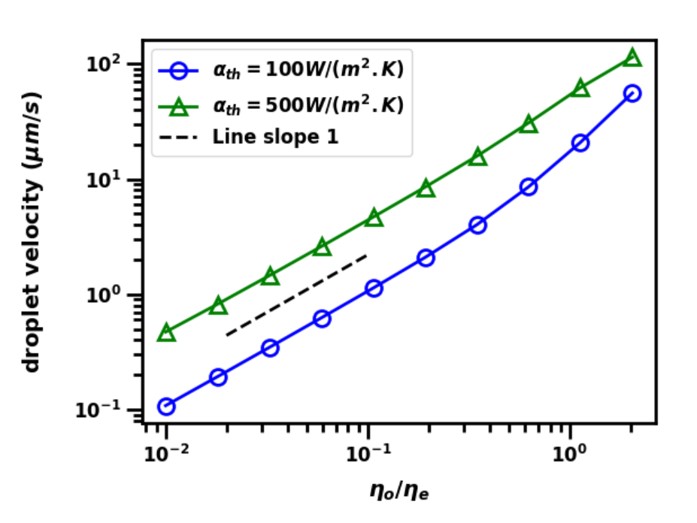}
\end{center}
\vspace{-15pt}
\caption{Droplet migration velocity versus the ratio of odd to even (shear) viscosity.
Finite-element numerical simulations of the full Navier-Stokes equations show that the velocity increases linearly with respect to odd viscosity in 
the proximity of experimentally determined odd viscosity values \cite{Soni2019}, where $\eta_o/\eta_e\sim 1/3$. Here we vary $\eta_o$ scaled by an $\eta_e$ that is determined from Table \ref{tab:table1} at \SI{38}{\celsius}.
\label{vel_v_eta} }
\vspace{-5pt}
\end{figure}

\begin{figure*}
\vspace{-5pt}
\begin{center}
\includegraphics[height=4.5in,width=7in]{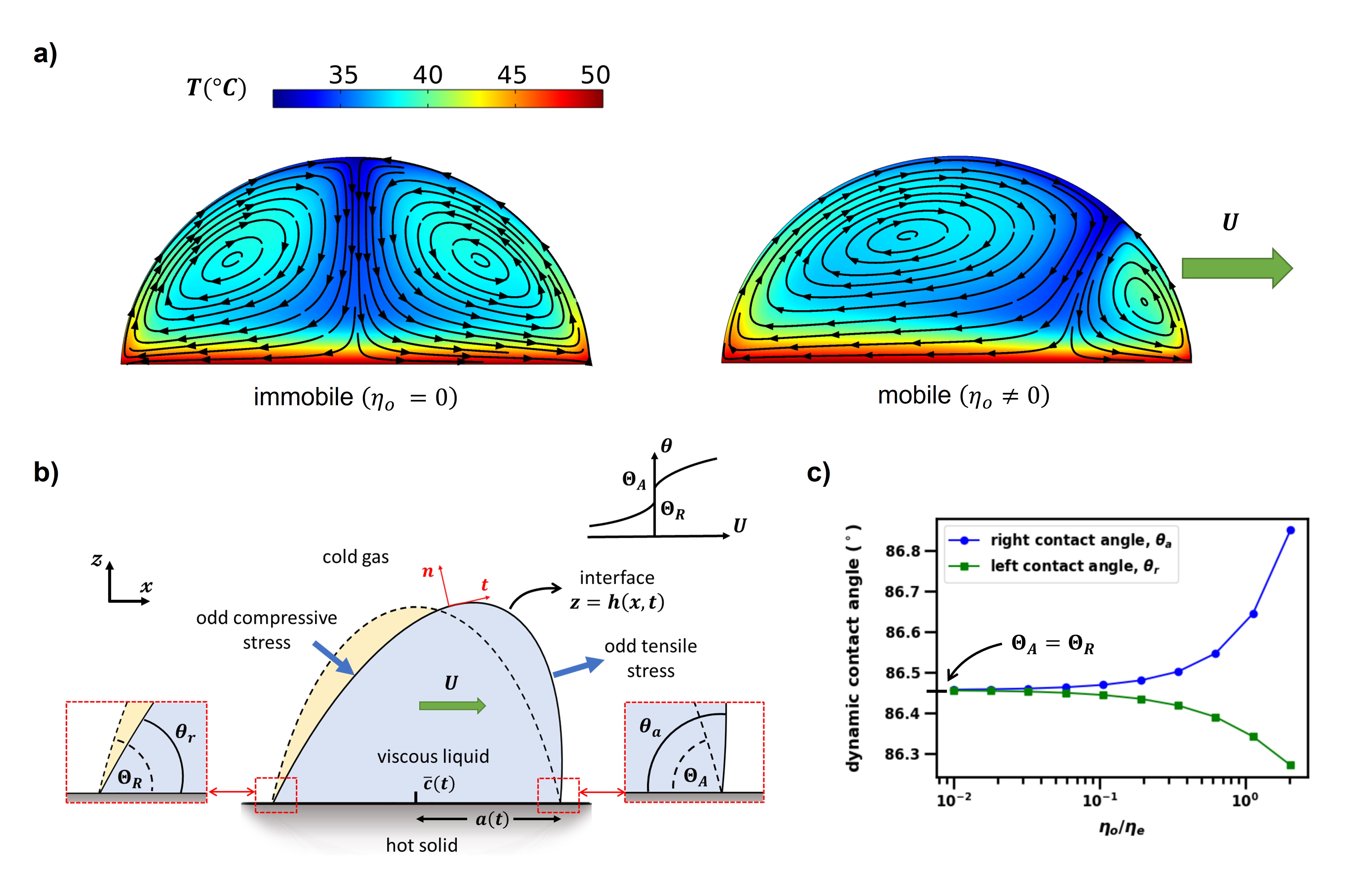}
\end{center}
\vspace{-5pt}
\caption{(a) Numerically determined thermocapillary droplet configurations in the absence and the presence
of odd viscosity. \textbf{Left:} Internal circulations of an opposite sense in a
thermocapillary droplet remaining immobile in the absence of odd viscosity \cite{Ehrhard1991,*Ehrhard1993}.  
\textbf{Right:} In the presence of odd viscosity the symmetry \rr{symmetry} breaks, leading to a left-right asymmetry in the cells and contact angles (cf. panel (c)), 
a commensurate droplet profile deformation and the onset of migration (cf. Fig. \ref{vel_v_eta}).
{ Compare panel (a) with the corresponding configurations of a \emph{three-dimensional} thermocapillary droplet displayed in Fig. \ref{odd_viscosity_droplet3}}. 
(b) Explanation of the odd viscosity-induced droplet migration effect in terms of contact angle asymmetry.
(b) \textbf{Top right}: Experiment-inspired hysteretic diagram of a single contact line moving with velocity $U$ vs.  dynamic contact angle $\theta(t)$ (cf. Eq.\rr{Ucl}). Contact lines move only when the dynamic contact angle $\theta(t)$ 
lies outside the interval $[ {\Theta}_R, {\Theta}_A]$, 
determined by the \emph{static} advancing and receding contact angles ${\Theta}_A$ and ${\Theta}_R$, respectively.
(b) \textbf{Main figure:}
Odd viscosity induces a compressive stress on the left cell and tensile on the right. 
The droplet tilts to the right and forces the right dynamic contact angle $\theta_a(t)$ to exceed its static
advancing counterpart $\Theta_A$ (right inset) and the left dynamic contact angle $\theta_r$ to lag behind the static receding contact angle $\Theta_R$ (left inset). 
This leads the droplet to migrate to the right with velocity $U$. 
$a(t)$ is the radius and $\bar{c}(t)$ is the middle point of the droplet. 
(c) right and left dynamic contact angles during droplet migration as a function of the viscosity ratio. 
Here we vary $\eta_o$ scaled by an $\eta_e$ that is determined from Table \ref{tab:table1} at \SI{38}{\celsius}.
\label{odd_viscosity_droplet} }
\vspace{-15pt}
\end{figure*}

{It is important to note here that the effect we develop describes how a stationary droplet with an 
effective temperature gradient \emph{normal} to the substrate will migrate because of odd viscosity. 
This effect differs from ones that employ temperature gradients lying \emph{horizontally} 
to the substrate and can give rise to liquid droplet migration even in the absence of odd viscosity \cite{Brochard1989, *Brzoska1993}.}

The numerically-determined dependence of migration velocity $U$ on odd viscosity, is displayed in Fig. \ref{vel_v_eta}
where $\eta_e$ is the even (shear) viscosity of the liquid. 
For instance, when $\eta_o/\eta_e\sim 1/3$ we obtain
\be \label{nest}
U \sim 5\; \mu \textrm{m}/\textrm{sec}. 
\ee

{To derive this result we considered \emph{temperature-dependent} material parameters} employed in 
representative advancing contact line experiments of a low volatility
ester, di-n-butyl phthalate \cite{Blake2019}, cf. Table \ref{tab:table1}, and a hydrophobic
substrate giving rise to a large static contact angle. We derived analogous results for other materials, 
such as silicon oils employed in spreading experiments \cite{Ehrhard1993}. 

A theoretical scaling law for the migration velocity can be established by following standard 
lubrication approximation arguments  (cf. \cite{Oron1997, *Davis2002,*Ehrhard1991,*Smith1995,*Aggarwal2023} and the SI) 
\be \label{Ucl0}
U \sim K \frac{\eta_o}{\eta_e}\frac{ \alpha_{th}a\Delta T}{2 k_{th}\gamma}\frac{d \gamma}{dT}
\ee
and was also numerically confirmed. 
Here, $\alpha_{th}$ is the 
heat transfer coefficient, $k_{th}$ is the liquid thermal conductivity satisfying $k_{th}\mathbf{n}\cdot\nabla T + \alpha_{th}(T-T_\infty)=0$, at the liquid-gas interface, $T_\infty$ is the ambient gas phase temperature, $a$ is the droplet radius, $\Delta T$ the temperature difference between the hot wall
and the cold air, 
$\gamma$ is surface tension and $\mathbf{n}$ is the normal, outward pointing, unit vector on the 
liquid-gas interface (cf. Fig. \ref{fig1}). 
$K$ is the mobility coefficient (units cm/sec, cf. Eq. \rr{Ucl}) which has to be determined by experiment. 
{Eq. \rr{Ucl0} was derived in the small $\Delta T$ limit where material parameters, other than
surface tension, do vary with temperature. To maintain accessibility, we tacitly assume in this article that the odd viscosity coefficient $\eta_o$ is a positive real number. Results with an odd viscosity coefficient of the opposite sign can then be obtained by making appropriate changes to our results.  }


Considering material values from Tables \ref{tab:table1} \& \ref{tab:table2} with 
$\Delta T=\SI{30}{\celsius}$, $T=\SI{38}{\celsius}$, $\alpha_{th} = 100 \textrm{W/(m}^2\cdot K)$ and 
$\eta_o/\eta_e = 1/3$,
we obtain from the scaling law \rr{Ucl0} that $U/K \sim 0.011$. A rough estimate of the migration velocity can be obtained by employing an order of magnitude assessment of $K\sim 10^{-3}$ m/sec from representative spreading experiments \cite{Ehrhard1993}. 
This gives
\be
U \sim 11\; \mu \textrm{m}/\textrm{sec}, 
\ee
which is in good order-of-magnitude agreement with the numerically-determined value \rr{nest}. 
{This result should only be followed as a guide
since currently there is only one experimental verfication for the presence of odd viscosity in a classical liquid
\cite{Soni2019}. }

%

Droplet migration is a consequence of symmetry-breaking.  In the absence of odd viscosity ($\eta_o \equiv 0$) the Navier-Stokes 
equations 
and boundary conditions are invariant with respect to the reflection symmetry
\be \label{symmetry}
 x \rightarrow -x, \quad u \rightarrow - u, 
\ee
where $u$ is the horizontal component of liquid velocity. This invariance is associated with the two symmetric internally-circulating cells of a thermocapillary droplet \cite{Ehrhard1991}, also depicted in the left droplet configuration of Figures \ref{fig1} and \ref{odd_viscosity_droplet}(a) (the latter is the numerically-determined 
droplet configuration). This invariance is also inherited by the contact angles: in the absence of odd viscosity
both left and right contact angles are equal to each other during spreading, retraction and equilibrium
as shown in the left droplet configuration of both Figures \ref{fig1} and \ref{odd_viscosity_droplet}(a).
On the other hand, in the presence of odd viscosity, this symmetry breaks in both the Navier-Stokes equations, 
and boundary conditions. This broken symmetry leads to a left-right asymmetry in the cells (cf. right configurations of Figures \ref{fig1} and \ref{odd_viscosity_droplet}(a), where the latter is numerically determined), unequal left and right (dynamic) contact angles, 
droplet tilting and migration. Migration is a general effect that arises every time 
this reflection symmetry breaks \footnote{A discussion of symmetry-breaking in the context of 
thermocapillary droplets is delegated to the Supplementary Information}.

A physical explanation of the odd viscosity-induced droplet migration effect developed in this Letter, is based on analyzing the stresses imparted on the liquid-gas interface 
by odd viscosity. 
The traction at the liquid-gas interface 
is \citep{Oron1997, *Davis2002}
\be  \label{freebc1}
\bm{\sigma}\mathbf{n} = \frac{\partial \gamma}{\partial s} \mathbf{t} + 2\kappa \gamma\mathbf{n}, 
\ee
where $\gamma$ is the (temperature-dependent) surface tension, $s$ is arc length along the interface, $\kappa$
its mean curvature and $\mathbf{t}$ and $\mathbf{n}$ are the unit tangent and (outward-pointing) normal vectors at the interface, see Fig. \ref{fig1}. The total stress incorporating
the effects of both even (shear) viscosity $\eta_e$ and odd viscosity $\eta_o$ is \cite{Avron1998,Khain2022} 
\be \label{seso}
\bm{\sigma} =-p\bm{I}+
 2\left( \begin{array}{cc}
 \eta_e & -\eta_o \\
 \eta_o & \eta_e
 \end{array} \right)
\mathcal{D}
\ee
where $\mathcal{D}_{ij} = \frac{1}{2}\left(\frac{\partial u_i}{\partial x_j} + \frac{\partial u_j}{\partial x_i}\right)$, $i=1,2$, $j=1,2$ is the rate-of-strain tensor.

%

\begin{table*}[t]
\caption{\label{tab:table1}%
Experimental data taken from \cite{Blake2019} and \cite{Haynes2016} for di-n-butyl phthalate (DBP). Curve
fitting of these material values is provided in the Supplementary Information. {Note that the temperature range for thermal conductivity differs from the range associated with other material parameters as they emanate
from two different references.}}
\begin{ruledtabular}
\begin{tabular}{llllll}
\textrm{Temp.} $\SI{}{\celsius}$&
$\eta_e$ ( mPa sec) &
$\gamma$ (mN $\textrm{m}^{-1}$) &
$\rho$ (kg $\textrm{m}^{-3}$) &
\textrm{Temp.} $\SI{}{\celsius}$&
$k_{th}$ (W $\textrm{m}^{-1}K^{-1}$) 
 \\
\colrule
15 &26.4 &35.0 &1049 &0& 0.139 \\
25 &16.7 &33.9 &1041&25&0.136\\
35 &11.2 &32.9 &1034&50&0.132\\
45 &8.1 &31.9 &1026&75&0.128\\
55 &6.6 &30.9 &1018&100& 0.125
\end{tabular}
\end{ruledtabular}
\end{table*}

The mechanical
behavior of interfaces between liquids is customarily described by the rate-of-strain tensor. For instance, $\mathbf{n} \mathcal{D} {\mathbf{n}}$
is the rate of extension, per unit length, of a material line element, which, in the current configuration, 
lies in the 
direction ${\mathbf{n}}$ \cite{Chadwick1976, *Truesdell1992}. It is also the deviatoric (zero-trace) part of
the normal stress at an interface. 
With respect to this description we provide below a qualitative interpretation
for the effect odd viscosity has on the droplet liquid-gas interface of Fig. \ref{fig1}. 
Employing \rr{freebc1} and \rr{seso}, the normal component of the rate-of-strain tensor $\mathcal{D}$ at 
the liquid-gas interface reads
\be \label{nDn}
{\mathbf{n}} \mathcal{D} {\mathbf{n}} = \frac{  -\eta_o\frac{\partial \gamma}{\partial s}+ \eta_e (p + 2\kappa \gamma )  }{2(\eta_e^2  + \eta_o^2)}. 
\ee 
Thus, odd viscosity gives rise to an extra normal stress proportional to $\eta_o$. It is 
compressive on the left liquid-gas interface (cf. Fig. \rr{fig1}) since $\frac{\partial \gamma}{\partial s} >0$ there, and tensile on the right, since $\frac{\partial \gamma}{\partial s} <0$ ($s$ is arc length along
the interface). This is the driving mechanism behind the droplet migration and 
is depicted in Fig. \ref{fig1}. 
The direction of the streamlines in Fig. \ref{odd_viscosity_droplet} is compatible with this interpretation.
This leads to a tank-tread interface motion, 
also compatible with the experiments of Dussan V and Davis \cite{Dussan1974a}. Note also that it is 
the term $\eta_o\frac{\partial \gamma}{\partial s}$ that breaks the reflection symmetry \rr{symmetry} in the 
boundary condition \rr{nDn}.

Symmetry between the contact angles also breaks in the presence of odd viscosity. 
While in the absence of odd viscosity the left
and right contact angles are equal, in the presence of odd viscosity they become mismatched and the 
droplet tilts with respect to its center axis (see the exaggerated configuration to the right of Fig. \ref{fig1}
and panel (b) in Fig. \ref{odd_viscosity_droplet}). 
This takes place when the aforementioned compressive and tensile stresses at the liquid-gas interface 
emanating from the term $\eta_o\frac{\partial \gamma}{\partial s}$ in Eq. \rr{nDn}, enable the contact lines to 
overcome hysteresis effects. In this case, the right dynamic contact angle $\theta_a$ exceeds the 
static advancing contact angle $\Theta_A$, (cf. right inset of Fig.~\ref{odd_viscosity_droplet} (b)) \emph{and} the left dynamic contact angle $\theta_r$ falls short of its receding static counterpart $\Theta_R$, (cf. left inset of Fig.~\ref{odd_viscosity_droplet} (b)). This leads to migration with a velocity $U$.

Surface roughness, chemical contamination and solutes may lead to contact line pinning whenever
the dynamic contact angle $\theta(t)$ lies within a finite interval $\Theta_R<\theta(t)<\Theta_A$ \cite{deGennes1985}. 
This is the phenomenon of contact-angle hysteresis (cf. upper cartoon in panel (b) of Fig. \ref{odd_viscosity_droplet}) experimentally documented by Dussan V \cite[Fig.~2]{Dussan1979}
relating contact angle with contact line velocity $U_{cl}$. In modeling this behavior, one finds that the contact
line moves with a velocity \cite{Oron1997}
\be \label{Ucl}
U_{cl} =  \pm K
\left\{  \begin{array}{cc}
(\theta - \Theta_A)^m, & \theta>\Theta_A,\\  
-(\Theta_R - \theta)^m, & \theta < \Theta_R
\end{array}
\right.
\ee
where $\theta$ is the dynamic contact angle (here denoted either as $\theta_a$ or $\theta_r$, cf. Fig.~\ref{odd_viscosity_droplet})
and the upper/lower sign is identified with the motion of the right/left contact line. $K$ is the mobility
coefficient appearing in the scaling law \rr{Ucl0}. 
Theoretical analyses \cite{Ehrhard1991,deGennes1985} and multiple spreading experiments from different groups 
\cite{Marsh1993,*Ehrhard1993,*Tanner1979, *Chen1988} established the consistency of the 
constitutive law \rr{Ucl} with experiment and determined values of the exponent $m$ according to the 
physical systems under consideration.

Odd viscosity-assisted migration of thermocapillary droplets is also expected to follow the law \rr{Ucl}. Neglecting hysteresis effects for simplicity, we determine numerically (cf. Fig. \ref{vel_v_eta}) that $m\sim1$, 
for odd viscosity values close to experiment ($\eta_o/\eta_e\sim 1/3$, \cite{Soni2019}). 
The value $m=1$ was employed in deriving the theoretical scaling law Eq. \rr{Ucl0}. 
Its derivation
follows general principles of the lubrication approximation 
and is delegated to the Supplementary Information.

\begin{table}[b]
\vspace{-10pt}
\caption{\label{tab:table2}%
Parameters employed in this article}
\begin{ruledtabular}
\begin{tabular}{lcl}
\textrm{Quantity}&
\textrm{Value}&
\textrm{Definition}\\
\colrule
$a$ (m)  & $1\times10^{-3}$& characteristic droplet radius\\
$T_0$  \SI{}{\celsius} &50 & solid substrate temperature\\
$T_\infty$  \SI{}{\celsius} &20 & gas phase temperature\\
$\Theta_A, \Theta_R$ degrees& 86.45&  static contact angles
\end{tabular}
\end{ruledtabular}
\end{table}

\begin{figure*}
\vspace{-5pt}
\begin{center}
\includegraphics[height=2.5in,width=7in]{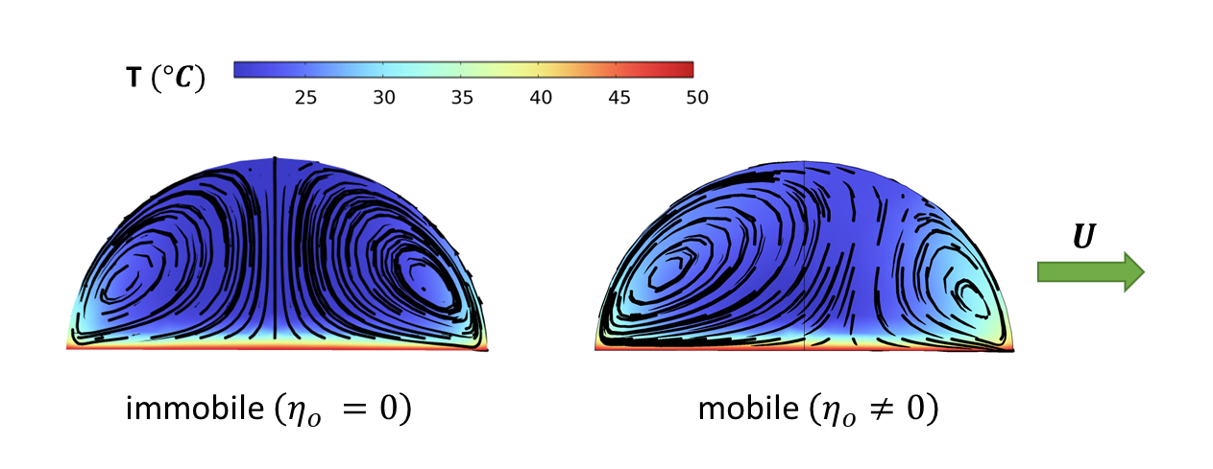}
\end{center}
\vspace{-5pt}
\caption{{Numerically-determined typical cross-sections of a \emph{three-dimensional} thermocapillary droplet  in the absence and in the presence
of odd viscosity. 
Compare with panel (a) of Fig. \ref{odd_viscosity_droplet} depicting
the corresponding states of a two-dimensional thermocapillary droplet. 
See Fig. S-I of the supplementary information for 
three-dimensional views taken from different viewing angles. 
\textbf{Left:} Counterrotating cells in a
thermocapillary droplet which remains immobile in the absence of odd viscosity. Observables in this
three-dimensional configuration such as contact angles, cell geometry and velocity field are symmetric with respect to the center axis. 
\textbf{Right:} In the presence of odd viscosity the axial symmetry is lost, leading to the onset of migration.
\label{odd_viscosity_droplet3}} }
\vspace{-5pt}
\end{figure*}

{In this Letter we employed a two-dimensional
geometry which permitted some analytical investigation of the odd viscosity-induced droplet migration effect. }Experimentally, two-dimensional droplet migration has 
been realized by placing oil films on vertical surfaces, for instance a wire,
and let it drop off in the form of a thin cylindrical stream \cite{Tanner1978, *Tanner1979}. 
This same two-dimensional
geometry was also employed theoretically to study spreading of droplets \cite{Ehrhard1991} and migration
under a horizontal temperature gradient \cite{Smith1995}
in the gravity or surface tension
dominated regime. 

{It is instructive to qualitatively compare our numerical results on the migration of 
a \emph{two-dimensional} thermocapillary droplet depicted in Fig. \ref{odd_viscosity_droplet},
with its
\emph{three-dimensional} counterpart. In Fig. \ref{odd_viscosity_droplet3} we display
typical cross sections of such three-dimensional droplets in the presence and in the absence of 
odd viscosity, to be compared with panel (a) of Fig. \ref{odd_viscosity_droplet} depicting
the corresponding states of a two-dimensional thermocapillary droplet. We witness the 
same symmetric recirculating counter-rotating cells in the absence of odd viscosity, where
the droplet necessarily remains immobile (left configuration of Fig. \ref{odd_viscosity_droplet3}
and compare with the left configuration of panel (a) in Fig. \ref{odd_viscosity_droplet}). 
In the presence of odd viscosity however, the symmetry of the three-dimensional droplet 
with respect to the center axis breaks leading to migration (right configuration of Fig. \ref{odd_viscosity_droplet3}
and compare with the right configuration of panel (a) in Fig. \ref{odd_viscosity_droplet}). 
In Fig. S-I of the supplementary information of this Letter we display different three dimensional
views (top view, 90 degree-elevated view) of the three-dimensional thermocapillary droplet 
whose cross-section is displayed in Fig. \ref{odd_viscosity_droplet3}. }

We thank the Department of Energy, Office of
Basic Energy Sciences for support under contract DE-FG02-08ER46539 and 
three anonymous referees for comments that improved the manuscript. 

%
%

\bibliographystyle{apsrev4-2}

\def\cprime{$'$}

\end{document}


\preprint{APS/123-QED}

\title{Thermocapillary migrating odd viscous droplets; Supplementary Information}
\author{A.Aggarwal}
\affiliation{Department of Materials Science \& Engineering, Robert R. McCormick School of Engineering and Applied Science, Northwestern University, Evanston IL 60208 USA and \\Center for Computation and Theory of Soft Materials, Northwestern University, Evanston IL 60208 USA
}
\author{E.Kirkinis}
\affiliation{Department of Materials Science \& Engineering, Robert R. McCormick School of Engineering and Applied Science, Northwestern University, Evanston IL 60208 USA and \\Center for Computation and Theory of Soft Materials, Northwestern University, Evanston IL 60208 USA
}
%
%
\author{M. Olvera de la Cruz}
\email{m-olvera@northwestern.edu}
\affiliation{Department of Materials Science \& Engineering, Robert R. McCormick School of Engineering and Applied Science, Northwestern University, Evanston IL 60208 USA and \\Center for Computation and Theory of Soft Materials, Northwestern University, Evanston IL 60208 USA
}

\date{\today}

\maketitle
\section{Results for the migration of a three-dimensional odd thermocapillary droplet}
\begin{figure*}
\vspace{-5pt}
\begin{center}
\includegraphics[height=7in,width=7in]{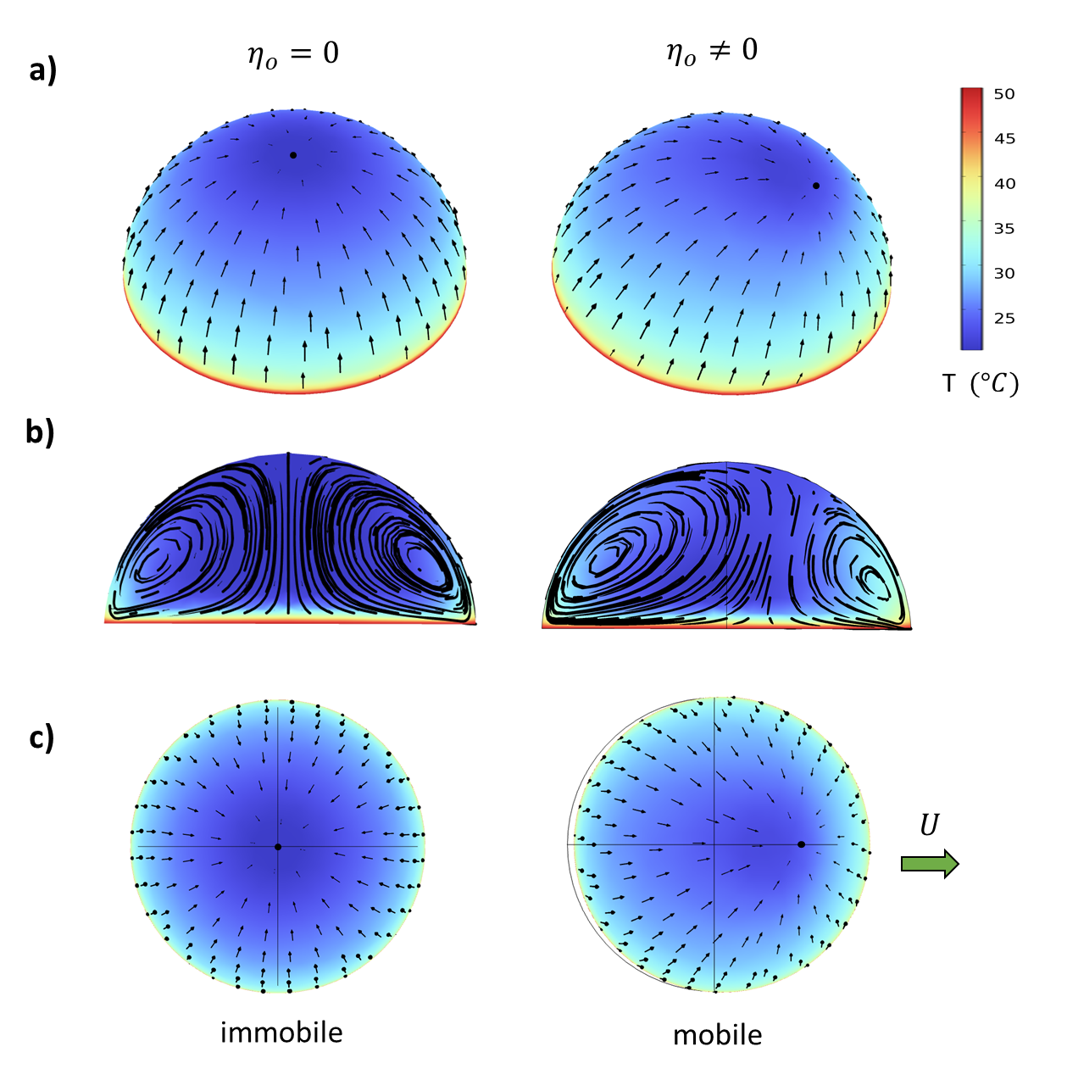}
\end{center}
\vspace{-5pt}
\caption{{Numerically determined thermocapillary droplet configurations in the absence (left column) and the presence (right column) 
of odd viscosity. (a) Left: in the absence of odd viscosity 
the thermocapillary effect leads to axisymmetric circulations. The arrows depict
the velocity field at the liquid-gas interface poining towards the summit. Right: In the presence of odd
viscosity, where the anisotropy axis is parallel to the solid substrate the axial symmetry of the contact
angles break leading to a recircularing cell deformation and the velocity field to point towards 
a point located off the geometric pole. This leads to droplet migrating to the right. (b) Cross section of the droplets displayed in panel (a). 
\textbf{Left:} Counterrotating cells in a
thermocapillary droplet which remains immobile in the absence of odd viscosity.  
\textbf{Right:} In the presence of odd viscosity the symmetry breaks, leading to a left-right asymmetry in the cells and the onset of migration.
(c) Top view of the three-dimensional droplet of panel (a). It is clear that the velocity field is axisymmetric
in the left (zero odd viscosity) configuration and asymmetric in the right (nonzero odd viscosity) configuration
leading to droplet migration.}
\label{odd_viscosity_droplet} }
\vspace{-15pt}
\end{figure*}

{In the main body of this article we have followed the two dimensional geometry that
is exclusively employed in the literature to computationally and theoretically study the migration of droplets. 
In this section we build on the theory developed in the main article 
and show that a \emph{three-dimensional} thermocapillary droplet composed of an odd viscous liquid
shows the same migration trend as in the two-dimensional case. To this end we employ the 
constitutive law \citep[\S13]{Landau1981} for the odd viscous stress
\be \label{sigma1}
\bm{\sigma}^o = \eta_o 
\left(\begin{array}{ccc}
-\left(\partial_x v + \partial_y u \right) &  \partial_x u  - \partial_y v & 0\\
 \partial_x u  - \partial_y v & \partial_x v + \partial_y u   & 0\\
0&0&0
\end{array}
\right).
\ee
which is similar to Eq (6) of the main article but here the liquid velocity is three dimensional
$(u,v,w)$, satisfies the incompressibility condition $\partial_x u + \partial_y v + \partial_zw=0$ and we take the $z$ axis to point out of the page as in panel (b) of Fig. \ref{odd_viscosity_droplet}. 

We numerically simulate the three-dimensional droplet by employing the same material parameters
as in the two-dimensional case with the heat transfer coefficient 
$\alpha_{th} = 500 \textrm{W/(m}^2\cdot K)$ and employing the same software package, as 
this is described in detail at the end of this SI. In the main article the qualitative 
comparison of panel (b) of Fig. \ref{odd_viscosity_droplet} with panel (a) of Fig. 3 of the main article
showed that the three-dimensional droplet follows the same trend as its two-dimensional counterpart.  
Panels (a) and (c) of Fig. \ref{odd_viscosity_droplet} also display the trend we would expect from
studying the droplet configurations of the main article. In the absence of odd viscosity (left column) 
a toroidally shaped cell encompases the center axis, so that the contact angles, cell and velocity field
are all axisymmetric. In the presence of odd viscosity (right column) the axial symmetry is lost and 
the droplet starts migrating to the right. 
}

\section{Governing equations and boundary conditions}
\subsection{Constitutive laws}
In an incompressible liquid endowed with odd viscosity the Cauchy stress tensor $\bm{\sigma}$
becomes \citep{Avron1998}
\be\label{totalstress}
\bm{\sigma} =-p\bm{I}+ \bm{\sigma}^e + \bm{\sigma}^o. 
\ee 
In two dimensions 
\be \label{evenstress}
\sigma^e_{ik} = \eta_e \left( \frac{\partial u_i}{\partial x_k} + \frac{\partial u_k}{\partial x_i}
\right) , \quad i,k=1,2,
\ee
is the standard deviatoric part of the Cauchy stress tensor with shear viscosity
$\eta_e$. 
$\bm{\sigma}^o$ is the odd part of the Cauchy stress tensor
\cite{Avron1998,Lapa2014}
\be  \label{oddstress}
\sigma^o_{ik} = - \eta_o (\delta_{i1}\delta_{k1} - \delta_{i3}\delta_{k3} ) 
\left( \frac{\partial u_1}{\partial x_3} +\frac{\partial u_3}{\partial x_1}  \right)
+ \eta_o \left( \delta_{i1}\delta_{k3} + \delta_{i3} \delta_{k1}    \right) 
\left( \frac{\partial u_1}{\partial x_1} - \frac{\partial u_3}{\partial x_3}\right),
\ee
where $\eta_o$ is the odd viscosity coefficient. 
%

Incorporating in \rr{totalstress} the contributions \rr{evenstress} and \rr{oddstress} we obtain a more
transparent form for the Cauchy stress tensor, cf. \cite{Kirkinis2023a}, 
\be \label{seso}
\bm{\sigma} =-p\bm{I}+
2 \left( \begin{array}{cc}
 \eta_e & -\eta_o \\
 \eta_o & \eta_e
 \end{array} \right)
\mathcal{D},
\ee
where $\mathcal{D} =  \frac{1}{2} \left(\frac{\partial u_i}{\partial x_j} + \frac{\partial u_j}{\partial x_i}\right)$ is the rate-of-strain tensor.

\subsection{Equations of motion and boundary conditions}

{The Navier-Stokes equations (for temperature-independent viscosity and density) become}
\begin{eqnarray}  \label{nsx}
\rho\left[ u_t + uu_x +wu_z\right] & = & -p_x  + \eta_e(u_{xx} + u_{zz}) - \eta_o(w_{xx} +  w_{zz}) , \\
\rho\left[ w_t + uw_x +ww_z\right] & = & -p_z + \eta_e (w_{xx} + w_{zz}) + \eta_o(u_{xx} + 
u_{zz}) - \rho g, \label{nsy}
\end{eqnarray}
where $(u,w)$ is the liquid velocity, $g$ the acceleration of gravity and $\rho$ the liquid density. 
At the liquid-gas interface $z = h(x,t)$, the shear and normal stresses are respectively
\be  \label{freebc1}
\mathbf{t} \bm{\sigma}\mathbf{n} = \frac{\partial \gamma}{\partial s} + \tau_0, \quad \textrm{and}
\quad \mathbf{n} \bm{\sigma}\mathbf{n} = 2\kappa \gamma + \Pi,  \quad \textrm{at}\quad
z = h(x,t),
\ee
where $\gamma$ is the (variable) surface tension, $s$ is arc length along the interface, $\kappa$
the mean curvature of the surface, $\tau_0$ is an external shear stress and $\Pi$ an external normal 
stress on the boundary \cite{Oron1997}. 

Boundary conditions at the liquid-solid interface are 
\be \label{solidbc1}
u = 0, \quad \textrm{and} \quad w=0, \quad \textrm{at} \quad z=0.
\ee

The liquid-gas interface is described by the function $z=h(x,t)$. The tangent and normal vectors to the interface are 
\be
\mathbf{t} = \frac{(1,h_x)}{\sqrt{1 + h^2_x}}, \quad \textrm{and}\quad \mathbf{n} = \frac{(-h_x,1)}{\sqrt{1 + h^2_x}}.
\ee
The components of stress tensor on the liquid-gas interface become
\be \label{tsn}
\mathbf{t} \bm{\sigma}\mathbf{n} = \frac{1}{1+h_x^2} (1,h_x)
\left( \begin{array}{cc}
\sigma_{11} & \sigma_{13} \\
\sigma_{31} & \sigma_{33}
\end{array}
\right)\left( \begin{array}{c}
-h_x \\ 1
\end{array} \right) = \frac{1}{1+h_x^2} \left[ \sigma_{13} - h_x^2 \sigma_{31} + h_x ( \sigma_{33} - \sigma_{11} ) \right],
\ee
and 
\be \label{nsn}
\mathbf{n} \bm{\sigma} \mathbf{n} = \frac{1}{1+h_x^2} (-h_x,1)
\left( \begin{array}{cc}
\sigma_{11} & \sigma_{13} \\
\sigma_{31} & \sigma_{33}
\end{array}
\right) 
\left( \begin{array}{c}
-h_x \\ 1
\end{array} \right) = \frac{1}{1+h_x^2} \left[ \sigma_{33} + h_x^2 \sigma_{11} - h_x ( \sigma_{13} + \sigma_{31} ) \right], 
\ee
so that
\be\label{freebc2a}
\frac{1}{1+h_x^2} \left[ \sigma_{13} - h_x^2 \sigma_{31} + h_x ( \sigma_{33} - \sigma_{11} ) \right]= \frac{\partial \gamma}{\partial s} + \tau_0,
\ee
and
\be
 \frac{1}{1+h_x^2} \left[ \sigma_{33} + h_x^2 \sigma_{11} - h_x ( \sigma_{13} + \sigma_{31} ) \right]=
2\kappa \gamma + \Pi, \label{freebc2b} 
\ee
at $z=h(x,t)$.  
The kinematic condition at $z=h(x,t)$ reads
\be \label{kinematic1}
w = h_t + u h_x.
\ee

%

\subsection{Normal and shear components of rate-of-strain tensor with respect to $\eta_o$}
Invoking the form of the Cauchy stress tensor \rr{seso},  
the
boundary conditions \rr{freebc1} at a free surface can be rewritten as 
\be \label{tsn2}
2\left(\eta_e\mathbf{t}-{\eta_o} \mathbf{n} \right)
\mathcal{D} \mathbf{n} = \frac{\partial \gamma}{\partial s} + \tau_0, \quad \textrm{at}\quad
z = h(x,t),
\ee
and 
\be \label{nsn2}
 -p + 2\left(\eta_e\mathbf{n}+{\eta_o}\mathbf{t} \right)
\mathcal{D} \mathbf{n}=2\kappa \gamma + \Pi,  \quad \textrm{at}\quad
z = h(x,t). 
\ee

Equations \rr{tsn2} and \rr{nsn2} can be considered as a system of two simultaneous equations for the
two unknowns $\mathbf{t} \mathcal{D} \mathbf{n} $ and $\mathbf{n} \mathcal{D} \mathbf{n} $.  
We can therefore rewrite each of the familiar components of the twice rate-of-strain tensor 
at an interface  in the form
\be \label{tDn}
\mathbf{t} \mathcal{D} \mathbf{n} = \frac{1}{2(\eta_e^2  + \eta_o^2)} \left[ \eta_e\left( \frac{\partial \gamma}{\partial s} + \tau_0 \right) + \eta_o (p + 2\kappa \gamma +\Pi )        \right]
\quad \textrm{at}\quad
z = h(x,t),
\ee
and 
\be \label{nDn}
\mathbf{n} \mathcal{D} \mathbf{n} = \frac{1}{2(\eta_e^2  + \eta_o^2)} \left[ -\eta_o\left( \frac{\partial \gamma}{\partial s} + \tau_0 \right) + \eta_e (p + 2\kappa \gamma     +\Pi )  \right]
\quad \textrm{at}\quad
z = h(x,t),
\ee

This is the main result of the first part of the SI. It expresses the fact that in the presence of odd viscosity 
the liquid will experience a 
normal stress-induced extra shear stress (the term multiplying $\eta_o$ in \rr{tDn})
and a shear stress-induced extra normal stress (the term multiplying $-\eta_o$ in \rr{nDn}) at the free surface.

\subsection{Thermocapillary flow\label{sec: thermo}}
In the system considered, there are temperature variations due to 
the thermocapillary effect and the applied temperature gradient. 
Conservation of energy 
{(for temperature-independent thermal conductivity and density)} reads
\be \label{energy1}
\rho c_p (T_t + \mathbf{u}\cdot \textrm{grad} T)  =  k_{th} \nabla^2 T
\ee
where $k_{th}$ is the liquid thermal conductivity, and $c_p$ the thermal expansion coefficient. 
In the geometry of our problem (cf. Fig.~\ref{fig1}),  Eq.~\rr{energy1} reduces to 
\be\label{energy1.1}
\rho c_p (T_t + uT_x + wT_z) = k_{th}(T_{xx} + T_{zz}), 
\ee
with boundary conditions 
\be \label{bc1}
T = T_0 \quad \textrm{at} \quad z=0 \quad \textrm{and} \quad 
k_{th} \mathbf{n}\cdot \nabla T + \alpha_{th} (T - T_\infty)=0, \quad \textrm{at} \quad
z = h(x,t),
\ee
where $T_0$ is the temperature of the rigid plane, $T_\infty$ 
the temperature of the gas phase and $\alpha_{th}$ the heat transfer coefficient. 

The surface tension is considered to be temperature dependent of the form
\be 
\gamma = \gamma_0  +  \frac{d\gamma}{dT} (T - T_0),
\ee
where $-\frac{d\gamma}{dT}$ is positive for common liquids \citep{Oron1997,Davis2002}.
We define 
$\Delta \gamma$ to be 
the variation of surface tension over the temperature domain 
$\Delta T = T_0 - T_\infty$
so that
\be \label{Deltasigma}
\Delta \gamma =\frac{d\gamma}{dT} \Delta T.
\ee

\subsection{Droplet migration}
\begin{figure}[t]
\begin{center}
\vspace{30pt}
\setlength{\unitlength}{1mm}
\begin{picture}(50,40)
\thicklines
\put(40,20){\vector(1,0){4}}
\put(40,22){$U$}

\qbezier
(0,5)(20,68)(40,5)
\put(-5,5){\line(1,0){50}}
\put(32,4){\vector(-1,0){10}} 
\put(30,4){\vector(1,0){10}}
\put(30,1){$a(t)$}
\put(16,6){$\bar c(t)$}
\put(21,4){\line(0,1){2}}
\thicklines
\qbezier
(35,15)(19,52)(21,15)
\qbezier
(5,15)(21,52)(19,15)
\put(19,15){\vector(0,-1){1}}
\put(21,15){\vector(0,-1){1}}

\put(50,49){$\displaystyle \theta $}
\put(62,36){$\displaystyle U $}
\put(52,36){\vector(0,1){15}}
\put(42,36){\vector(1,0){20}}  
\thicklines

\qbezier(52,46)(53,48)(60,50)
\qbezier(44,38)(51,40)(52,42)
\put(47,45){$\displaystyle {\Theta}_A $}
\put(52,40){$\displaystyle {\Theta}_R $}

\put(9,10){viscous liquid}
\put(-6,46){cold gas}
\put(10,0){hot solid}
\thicklines
\put(25,36){interface}
\put(28,33){$\displaystyle z=h(x,t)$}
\thicklines
\qbezier
(2,12)(7,10)(8,5)
\put(0,5){\line(1,4){4}}
\put(3,6){$\displaystyle \theta_r $}
\put(-1,2){$\displaystyle c_r(t) $}

\qbezier
(38,12)(34,10)(32,5)
\put(40,5){\line(-1,4){4}}
\put(35,6){$\displaystyle \theta_a $}
\put(39,2){$\displaystyle c_a(t) $}

\put(0,30){\vector(2,-1){8}}
\put(-21,31){odd compressive stress}

\put(31,28){\vector(2,1){8}}
\put(32,26){odd tensile stress}


\put(-19,8){\line(1,0){12}}
\put(-19,8){\line(1,4){3}}
\put(-19,8){\line(1,2){3}}
\qbezier(-17,12)(-15,11)(-14,8)
\qbezier(-17,18)(-10,17)(-8,8)
\put(-18,8.5){$\displaystyle \theta_r $}
\put(-15,11.5){$\displaystyle \Theta_R $}
\put(-20,7){\line(1,0){14}}
\put(-20,22){\line(1,0){14}}
\put(-20,7){\line(0,1){15}}
\put(-6,7){\line(0,1){15}}
\put(-5.2,7){\vector(3,-1){5}}
%

\put(48,8){\line(1,0){11}}
\put(59,8){\line(-1,4){3}}
\put(59,8){\line(-1,2){3}}
\qbezier(56.7,13)(53,12)(52,8)
\qbezier(57,18)(50,17)(48,8)
\put(53.7,8.8){$\displaystyle \Theta_A $}
\put(50.5,11.5){$\displaystyle \theta_a $}
%
\put(46,7){\line(1,0){14}}
\put(46,22){\line(1,0){14}}
\put(46,7){\line(0,1){15}}
\put(60,7){\line(0,1){15}}
\put(46,7){\vector(-3,-1){5}}

\put(-15,44){$\displaystyle z $}
\put(-2,36){$\displaystyle x $}
\put(-12,36){\vector(0,1){10}}
\put(-12,36){\vector(1,0){10}}  

\put(8,40){$\displaystyle \mathbf{n} $}
\put(20,40){$\displaystyle \mathbf{t} $}
\put(14,34){\vector(1,1){6}}
\put(14,34){\vector(-1,1){6}} 
\thicklines
\end{picture}
\caption{
\textbf{Top right}: Experiment-inspired hysteretic diagram of a single contact line moving with velocity $U_{\textrm{cl}}$ vs.  dynamic contact angle $\theta(t)$ (cf. Eq.\rr{Ucl} and \citep{Dussan1979}). Contact lines move only when the dynamic contact angle $\theta(t)$ 
lies outside the interval $[ {\Theta}_R, {\Theta}_A]$, 
determined by the \emph{static} advancing and receding contact angles ${\Theta}_A$ and ${\Theta}_R$, respectively.
\textbf{Main figure:}
Droplet configuration at the onset of migration:
the viscous liquid droplet on a hot solid substrate surrounded by a colder ambient gas phase. 
The thermocapillary effect
creates in the droplet two identical cells with opposite sense of circulation \cite{Ehrhard1991}. 
Odd viscosity induces a compressive stress on the left cell and tensile on the right. 
The droplet tilts to the right and forces the right dynamic contact angle $\theta_a(t)$ to exceed its static
advancing counterpart $\Theta_A$ (right inset) and the left dynamic contact angle $\theta_r$ to lag behind the static receding contact angle $\Theta_R$ (left inset). 
This leads the droplet to migrate to the right with velocity $U$ according to the experiments of Dussan V
\cite{Dussan1979} (cf. the upper right hysteretic diagram). 
\label{fig1} }
\end{center}
\vspace{-10pt}
\end{figure}
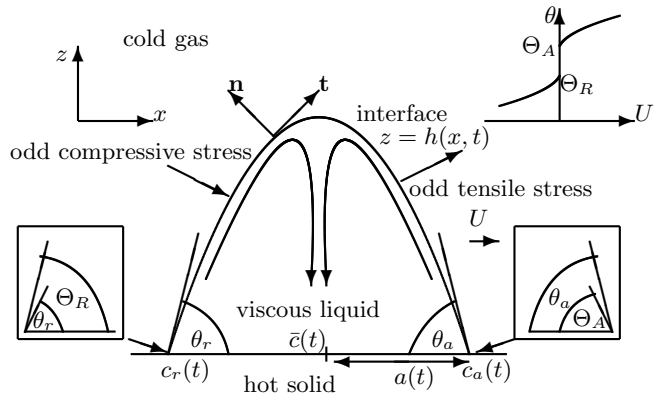

The problem of determining the motion of a droplet reduces to 
solving an evolution equation for the liquid-gas interface $z=h(x,t)$ with boundary conditions
\be \label{bcmob0}
h(c_a(t),t) = h(c_r(t),t) =0, \quad \int_{c_r(t)}^{c_a(t)} h(x,t)dx = V_0,
\ee
where 
\be \label{cathetaa}
c_a(t) \quad \textrm{and}\quad \theta_a(t),\quad \quad c_r(t) \quad \textrm{and}\quad \theta_r(t)
\ee
is the $x$-location of the contact line and dynamic contact angle at the right and the left contact lines
of the droplet, respectively (cf. Fig. \ref{fig1}) {and $V_0$ denotes the constant area of the two-dimensional
droplet}. 
Employing the nondimensional units introduced in \cite{Smith1995, Aggarwal2023} (which are not 
repeated here to avoid possible duplication), we adopt the mobility law 
\be \label{mob0}
\frac{dc_a}{dt} = \left\{ 
\begin{array}{cc}
(\theta_a-\Theta_A)^m, &  \theta_a>\Theta_A\\
0, & \Theta_R\leq \theta_a\leq \Theta_A\\
-(\Theta_R - \theta_a)^m& \theta_a<\Theta_R
\end{array} \right.
\quad
\textrm{and}
\quad
\frac{dc_r}{dt} = \left\{ 
\begin{array}{cc}
-(\theta_r-\Theta_A)^m, &  \theta_r>\Theta_A\\
0, & \Theta_R\leq \theta_r\leq \Theta_A\\
(\Theta_R - \theta_r)^m& \theta_r<\Theta_R
\end{array} \right.
\ee
where $\theta_a(t) = -h_x(c_a(t))$ and $\theta_r(t) = h_x(c_r(t))$ (cf. Fig. \ref{fig1})
and 
\be  \label{barthetaa}
\Theta_A\quad \textrm{and}\quad \Theta_R
\ee
are the advancing and receding \emph{static} contact angles, respectively (cf. upper right corner
of Fig. \ref{fig1}) { and we tacitly assumed that angles are small}.

\section{\label{sec: symmetry}Droplet migration is induced by symmetry-breaking}

\begin{figure}
\vspace{20pt}
\begin{center}
\setlength{\unitlength}{1.2mm}
\begin{picture}(90,30)
\thicklines

\put(10,7){\line(1,0){25}}
\qbezier(12,7)(23,50)(34,7)
\put(12,4){$u(-x,z) = -u(x,z)$}


\put(18,30){$\eta_o \equiv 0$}
\put(15,8){viscous liquid}
\put(30,25){gas}

\qbezier
(31,12)(24,40)(24,12)
\qbezier
(15,12)(22,40)(22,12)
\put(24,12){\vector(0,-1){1}}
\put(22,12){\vector(0,-1){1}}

\put(50,7){\line(1,0){25}}
\qbezier(52,7)(73,50)(74,7)
\put(12,4){$u(-x,z) = -u(x,z)$}

\qbezier
(72,12)(70,38)(68,12)
\qbezier
(56,12)(72,40)(66,12)
\put(68,12){\vector(0,-1){1}}
\put(66,12){\vector(0,-1){1}}

\put(52,4){symmetry-breaking}
\put(62,30){$\eta_o \neq 0$}
\put(57,8){viscous liquid}
\put(58,25){gas}

\end{picture}
\caption{Symmetries of the equations of motion for a thermocapillary droplet (left) in the absence of odd viscosity and symmetry breaking in a thermocapillary droplet in the presence of odd viscosity (right).
In the former case the Navier-Stokes equations, boundary conditions and contact angles are invariant
with respect to the transformation $x\rightarrow -x, u\rightarrow - u$. This gives rise to two recirculating
cells inside the droplet \cite{Ehrhard1991}. In the latter case odd viscosity 
breaks all former symmetries. This leads the droplet to tilt and migrate to the right. 
\label{fig3} }
\end{center}
\vspace{0pt}
\end{figure}
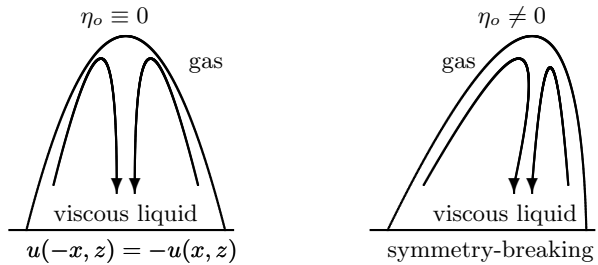

In this article we consider the case displayed to the right of Fig.\ref{fig3}. 
A droplet in the presence of odd viscosity breaks the reflection symmetry $x\rightarrow -x, u\rightarrow - u$, 
in the equations of motion \rr{nsx} and \rr{nsy}
and in the stress boundary conditions.
The loss of symmetry in the above system becomes evident in the interfacial boundary conditions
written with respect to the rate-of-strain tensor as was done in the main body of the Letter.  
Employing the general expressions \rr{tDn} and \rr{nDn}, the normal component of the rate-of-strain tensor $\mathcal{D}$ at $
z = h(x,t)$ 
\be \label{nDn1}
\mathbf{n} \mathcal{D} \mathbf{n} = \frac{  -\eta_o\frac{\partial \gamma}{\partial s}+ \eta_e (p + 2\kappa \gamma )  }{2(\eta_e^2  + \eta_o^2)}. 
\ee 
Thus, odd viscosity gives rise to an extra normal shear stress proportional to $\eta_o$. It is 
compressive on the left liquid-gas interface (cf. Fig. \rr{fig1}) since $\frac{\partial \gamma}{\partial s} >0$ there and tensile on the right, since $\frac{\partial \gamma}{\partial s} <0$ there, ($s$ is arc length along
the interface). This is the driving mechanism behind the droplet migration and 
is depicted in Fig. \ref{fig1}. Notice that this term breaks the symmetry
$x\rightarrow -x, u\rightarrow - u$ that existed in its absence. 

Likewise, in the shear component of the rate-of-strain tensor on the liquid-gas interface $
z = h(x,t)$ 
\be \label{tDn1}
\mathbf{t} \mathcal{D} \mathbf{n} = \frac{ \eta_e \frac{\partial \gamma}{\partial s}  + \eta_o (p + 2\kappa \gamma  )}{2(\eta_e^2  + \eta_o^2)}, 
\ee
the odd viscosity term breaks the $x\rightarrow -x, u\rightarrow - u$ symmetry that existed in its 
absence. In the context of the present problem the symmetry-breaking terms in \rr{tDn} lead
to a renormalization of the thermocapillary effect. In other cases however (cf. \cite{Soni2019})
the combination of \rr{nDn1} with \rr{tDn1} may lead to additional effects such as unidirectional
wave propagation. 

{The symmetry breaking mechanism of the contact lines was explained in detail for the case of 
a droplet migrating under the action 
of a magnetic torque (in the absence of odd viscosity) \cite{Aggarwal2023}. The analysis carried-out
in Eq. (7.3)-(7.6) and Fig. 14 of \cite{Aggarwal2023} can be employed here as well by making
the substitution
\be
N \rightarrow \eta M
\ee
where $N$ is the magnetic torque defined in \cite{Aggarwal2023}.

The magnetic torque-actuated \emph{isothermal} droplet (with no odd viscosity and no thermocapillarity) \cite{Aggarwal2023}
is driven by the motion of the liquid gas interface due to the collective rotation of the ferroparticles
suspended in the liquid and gives rise to a \emph{single} circulating cell. This differs
significantly from the \emph{non-isothermal} effect considered here where the thermocapillary effect
produces two cells with opposite sense of circulation and the droplet is set into motion purely by
the presence of odd viscosity. 

There are similarities however. In both the magnetic torque paper \cite{Aggarwal2023}
and in the present contribution we employ the same constitutive law for the droplet 
motion (Eq. \rr{mob0}), giving rise to the same symmetry-breaking mechanism based 
on the asymmetry of the contact angles. 
}

\begin{table*}[t]
\caption{\label{tab:table1}%
Experimental data taken from \cite{Blake2019} and \cite{Haynes2016} for di-n-butyl phthalate (DBP), valid
between $\SI{15}{\celsius}$ and $\SI{55}{\celsius}$ and curve-fitted.}
\begin{ruledtabular}
\begin{tabular}{lcl}
\textrm{Quantity}&
\textrm{Value}&
\textrm{Definition}\\
\colrule
$\rho$ ( $\textrm{kg}\:\textrm{m}^{-3} $)    & $\frac{1}{80000} T^{4}-\frac{11}{6000} T^{3}+\frac{151}{1600} T^{2}-\frac{661}{240} T +\frac{137553}{128}$        & density \\
$\eta_e$ ( mPa$\cdot$ sec)    & $4.166666667\times 10^{-6} T^{4}- 0.0008T^{3}+ 0.06495833333 T^{2}- 2.73T + 55.2234375$        & shear (or even) viscosity \\
$a$ (m)  & $1\times10^{-3}$& characteristic droplet radius\\
$\gamma$ (mN $\textrm{m}^{-1}$) &$4.16666\times 10^{-7} T^{4}- 0.000066666 T^{3}+ 0.003895833 T^{2}- 0.198333 T + 37.3023$& surface tension \\
$k_{th}$ (W $\textrm{m}^{-1}K^{-1}$) &$1.066666\times 10^{-8} T^{3}- 1.6\times 10^{-6} T^{2}- 0.00008666T + 0.139$& thermal conductivity  \\
$T_0$  \SI{}{\celsius} &50 & solid substrate temperature\\
$T_\infty$  \SI{}{\celsius} &20 & gas phase temperature\\
$\Theta_A, \Theta_R$ degrees& 86.45&  static contact angles
\end{tabular}
\end{ruledtabular}
\end{table*}

\section{\label{generalF}General expression for migration velocity in the perturbative regime}
Broken symmetry and a general expression for the migration velocity 
depend on the constitutive law \rr{mob0} and is thus identical to the migration of droplets
induced by different mechanisms \cite[Appendix B]{Aggarwal2023}. Thus, to avoid 
possible duplication here, as in the previous section, \emph{we have} to state only the final results. 

In the presence of odd viscosity, the left-right symmetry of contact angles breaks
\be \label{F}
\theta_a= F_0 +\eta M F_1, \quad \textrm{and} 
\quad
\theta_r= F_0 - \eta MF_1. 
\ee
The explicit form of $F_0$ and $F_1$ is not needed 
here {(the expression $F_0$ in \rr{F} depends on $M$, see Eq. \rr{thetara} below).}
Employing the arguments on contact angles stated in the Appendix of \cite{Aggarwal2023}, 
the migration velocity is then given by 
\be \label{UAR}
U = (\theta_a - \Theta_A)^m =\left[ F_0 + \eta M F_1 - \Theta_A\right]^m = \left[\eta M  F_1 - \frac{\Theta_A-\Theta_R}{2}\right]^m.
\ee
In the absence of hysteresis effects ($\Theta_A = \Theta_R \neq 0$) the migration velocity reduces to 
\be \label{U}
U = \left(\eta M F_1\right)^m. 
\ee

\section{Quantitative description of migration velocity}
To gain a quantitative insight into the odd viscosity droplet migration effect, we consider the problem of migration by employing the existing formulation
of droplet spreading and migration developed by Ehrhard \& Davis \cite{Ehrhard1991} and Smith \cite{Smith1995}, based on the lubrication approximation, 
and proceed on an \emph{ad-hoc} basis. 
The acquired closed form droplet migration velocity $U$ agrees with our numerical simulations, as this was established in
the main section of this Letter (for instance, by comparing Eq. (2) of the Letter with the Letter's Fig. 2). To this end, and starting from equation \rr{nDn} we consider the 
following line of thought
\be
\frac{\partial \gamma}{\partial s}\sim \frac{d \gamma}{dT}\frac{dT}{dh}h_x = 
\frac{\alpha_{th} \Delta T }{k_{th}} \frac{d\gamma}{dT} h_x
\ee
where we employed the temperature profile $T(h) = \frac{\Delta T}{1 + Bi h/h_0} + T_\infty$
and $Bi = \frac{\alpha_{th} h_0}{k_{th}}$
\cite{Ehrhard1991}, $h_0$ is a characteristic size for the droplet and the arc length has been replaced
by the horizontal coordinate (strictly, valid only in the long wavelength approximation). $T_\infty$ is the 
ambient gas phase temperature {and we employed the small Biot number limit which is characteristic for
most liquids \cite{Oron1997}.  }

Consider a viscous droplet, incorporating the effects of odd viscosity, sitting on a heated solid substrate and 
surrounded by a colder ambient gas phase and denote its liquid-gas interface by $z=h(x,t)$ (cf.~Fig.~\ref{fig1}).    
Conservation of momentum and energy
leads to velocity profiles derived by Ehrhard and Davis \cite[Eq.~(4.11p)]{Ehrhard1991}
with an amended pressure gradient due to the term $\eta_o\frac{\partial \gamma}{\partial s}$
appearing in the normal stress boundary condition \rr{nDn1}. 
In the quasistatic limit of small Biot and capillary numbers 
the dynamics is determined by the balance of pressure and thermocapillary boundary
motion
\be\label{bal2}
-\frac{1}{3}h^3 \bar p_x + \frac{1}{2}h^2\sigma_x =0,
\ee
where 
$-\bar p_x=\left[ h_{xx} - \eta {M h_x}
\right]_x$ and $\sigma_x = Mh_x$ by employing the dimensionless units developed in \cite{Ehrhard1991}. 
$M$ is the product
of Marangoni number (measures the ratio of the temperature gradient to the mobility of the contact line),
Biot number (measures the ratio of 
the rate of heat transfer from liquid to gas, to liquid thermal conductivity) and capillary number (measures 
the ratio of contact line mobility to surface tension) and is given by 
\be
M =  \frac{ \alpha_{th}a\Delta T}{\theta_0\gamma k_{th}}\left|\frac{d \gamma}{dT}\right|
\ee
where $\theta_0$ is a small angle which, however, drops out when dimensional units are used, 
$|\cdot|$ denotes the absolute value of a real number and $a$ is the droplet radius.  

The effect discussed in this Letter arises due to the term 
$\eta M h_x$ in the pressure gradient, where $\eta$ is the ratio of odd viscosity to its (even) standard
counterpart.
When the plate is heated ($M>0$) there is a compressive pressure on the left cell
(clockwise circulating) since $h_x>0$ there, and a tensile pressure on the 
right cell (anticlockwise circulating) where $h_x<0$. 
When the plate is cooled, ($M<0$) the circulation patterns and pressure directions just described are 
reversed.

Reverting to a frame of reference whose origin $\bar{c}(t)$ lies at the mid-point of the droplet base and denoting its spreading
radius by $a(t)$ \cite{Smith1995}, (cf.~Fig.~\ref{fig1}) we can solve Eq.~\rr{bal2} considering that the droplet height vanishes 
at the two contact angles
$
h(\pm a,t) = 0, $
and its volume is a constant
$ \int_{-a}^{a} h(x,t) dx = 1 
$
{(we employ dimensionless units in this calculation that have been defined in \cite{Smith1995} and \cite{Aggarwal2023}. The radius $a$ was nondimensionalized by its counterpart $a_0$
attained during migration).}
We expand all fields in a perturbation series
$
h = h_0  + M h_1 +..., 
$
and in the absence of gravity
we obtain the following liquid-gas interface (dimensionless) profile
\begin{widetext}
\be \label{h01}
h(x,t) \sim \frac{3}{4} \left\{ \frac{a^2 - x^2}{a^3} +M \left[\frac{\eta x (a^2 - x^2)}{3a^3} + 
2(a^2 + x^2)\ln(2a) - \frac{4}{3}(a^2  - x^2) - (a-x)^2 \ln(a-x) - (a+x)^2 \ln(a+x) \right] \right\}.
 \ee
\end{widetext}
%

Notice the asymmetry induced on the liquid-gas interface profile \rr{h01} by odd viscosity. Since
$|x|\leq a$, the profile is elevated to the right of the droplet and is suppressed to the left. When $\eta\equiv 0$ one recovers the spreading results of Ehrhard and Davis \cite[Eq.~(7.2p)]{Ehrhard1991} where
two identical cells with opposite sense of circulation are formed in the interior of the droplet (cf.~Fig. 1 and 2
of the main part of this Letter). {Also note that Eq. \rr{h01} was also derived by Smith \cite{Smith1995}
with the exception of the term multiplying the coefficient $\eta$. }

Employing \rr{h01}, the left and right dynamic contact angles $\theta_r = h_x(-a)$ and $\theta_a = -h_x(a)$ are, respectively, 
\be \label{thetara}
\theta_{a,r} = \frac{3}{2a^2} (1 - \frac{1}{3}Ma^3 \pm \frac{1}{3} \eta M a).
\ee

Surface roughness, chemical contaminations and solutes may lead to contact line pinning whenever
the contact angle $\theta$ lies within a finite interval $\Theta_R<\theta<\Theta_A$ \cite{deGennes1985}. 
This is the phenomenon of contact-angle hysteresis and has been experimentally documented by Dussan V \cite[Fig.~2]{Dussan1979}
relating contact angle with contact line velocity $U_{cl}$. In modeling this behavior, one finds that the contact
line moves with a velocity \cite{Oron1997}
\be \label{Ucl}
U_{cl} \sim \pm
\left\{  \begin{array}{cc}
(\theta - \Theta_A)^m, & \theta>\Theta_A,\\  
-(\Theta_R - \theta)^m, & \theta < \Theta_R
\end{array}
\right.
\ee
where $\theta$ is the dynamic contact angle (here denoted either as $\theta_a$ or $\theta_r$, cf. Fig.~\ref{fig1})
and the upper/lower sign is identified with the motion of the right/left contact line (Eq. \rr{Ucl} is a 
simplified version of the exact formula \rr{mob0}).  
Theoretical analyses \cite{Ehrhard1991,deGennes1985} and multiple experiments from different groups 
\cite{Marsh1993,*Ehrhard1993,*Tanner1979, *Chen1988} determine the value of the exponent $m$. 
In this Letter we determine that $m\sim 1$ employing numerical simulations. 

The motion of the droplet is composed of a transient and a steady-state configuration. Contact angle hysteresis endows the 
transient with diverse behavior:
spreading, pinning and migration where left and right contact line velocities are unequal.
These cases, exhaustively classified by Smith \cite{Smith1995} for a different problem, 
are applicable here as well. For instance, Eq.~\rr{thetara} shows that the droplet may spread
but the right-hand contact angle spreading will always be faster.

We are here interested on the steady-state configurations
attained by the migrating droplet. When the droplet has reached  its steady-state migration shape and velocity, 
Eq.~\rr{Ucl} implies that the dynamic contact
angles are related by (cf.~Fig.~\ref{fig1} for the definition of the angles)
\be
\theta_a - \Theta_A = \Theta_R - \theta_r, \quad \textrm{and} \quad \theta_a-\Theta_A>0. 
\ee
Employing Eq. \rr{thetara}, the first of the above expressions leads to 
$
\theta_a+\theta_r = \frac{3}{a^2}(1 - \frac{1}{3}Ma^3) = \Theta_A + \Theta_R, \quad
$
and the second to the inequality 
$
0< \frac{1}{2}(\Theta_A - \Theta_R) < \frac{\eta M}{2R}. 
$
Eq.~\rr{Ucl} leads to the migration (dimensionless) velocity
\be
U = (\theta_a - \Theta_A)^m = \left[ \frac{\eta M}{2a} - \frac{\Theta_A-\Theta_R}{2}\right]^m,
\ee
(applicable for a general $m$, see the discussion is section \ref{generalF}. 
In this article we consider migration of droplets in the absence of hysteresis ($\Theta_A = \Theta_R \neq 0$) and set $m=1$, thus, the migration velocity reduces to 
\be \label{UM2a}
U = \frac{\eta M}{2a}
\ee
which is the dimensionless counterpart of Eq. (2) of the Letter. 
Thus, droplet migration on a vertical temperature gradient is a purely odd viscosity-induced effect. 

The dimensional counterpart of Eq. \rr{UM2a} is given by (this is Eq.(2) of the Letter)
\be \label{Ucl0}
U \sim K \frac{\eta_o}{\eta_e}\frac{ \alpha_{th}a\Delta T}{2 k_{th}\gamma}\frac{d \gamma}{dT}.
\ee
Here, $\alpha_{th}$ is the 
heat transfer coefficient, $k_{th}$ is the liquid thermal conductivity satisfying $-k_{th}\mathbf{n} \cdot\nabla T + \alpha_{th}(T-T_\infty)=0$, at the liquid-gas interface, $T_\infty$ is the ambient gas phase temperature, $a$ is the droplet radius, $\Delta T$ the temperature difference between the hot wall
and the cold air, 
and $\gamma_0$ is the surface tension at wall 
temperature. $K$ is the mobility coefficient (units cm/sec). 
We have performed extensive numerical checks by varying \emph{all} the parameters appearing in 
Eq. \rr{Ucl0} and verified that they agree well in the vicinity of the currently available experimental 
values for the viscosity ratio $\eta_o/\eta_e\sim 1/3$ and that the exponent $m$ remains equal
to $1$.

This Letter paves the way to utilize odd viscosity in manipulating structures that incorporate
recirculating cell patterns. Interfacial driving mechanisms are particularly important in microfluidic devices
due to their large surface-to-volume ratio. Current interest lies on discrete or 
continuous microscopic control of small-scale systems and applications vary 
from clinical and forensic analysis to semiconductor devices and environmental 
monitoring \cite{Darhuber2005}.

\section{Numerical method}
We use a commercial finite element software, COMSOL\cite{comsol2022comsol} to model the odd viscous droplets studied in this letter. COMSOL uses the \textit{moving mesh} method defined within the framework of Arbitrary Lagrangian-Eulerian (ALE) formulation \cite{donea2004arbitrary} to define a deformable computational domain for the fluid (droplet). The geometric surface of the fluid domain acts as the free surface that separates the fluid from the surroundings. The interfacial effects due to the surrounding gas phase are modeled via surface tension (or any other boundary forces) applied directly on the fluid surface as boundary conditions. The advantage of using the moving mesh method is that only the fluid domain is explicitly defined. This leads to a substantial improvement in performance compared to other commonly used interface capturing methods such as the level set \cite{sethian2003level} or volume of fluid \cite{hirt1981volume} methods because we only solve for the fluid domain and ignore the gas domain. Additionally, the moving mesh method provides a much sharper and accurate interface because it is modeled as a geometric surface. However, it has the limitation of not being able to handle topological changes, such as a droplet splitting into two. Fortunately, this limitation is not relevant for the droplet system we study here.

\subsection{Modeling the equation of fluid motion}
We use COMSOL's in-built \textit{laminar flow} and \textit{heat transfer in fluids} modules to model the system. However, COMSOL assumes the viscosity to be a scalar field. In order to account for the odd stress $\boldsymbol{\sigma^o}$ in the model, we need to look at the weak formulation of governing equations and input the extra terms manually.

The governing equations for incompressible fluid motion are 

\begin{equation} \label{eq:NS-eom}
    \rho (\partial_t \boldsymbol{u} + (\boldsymbol{u}\cdot \nabla)\boldsymbol{u}) = -\nabla p +\nabla \cdot \boldsymbol{\sigma} + \boldsymbol{f}
\end{equation}
\begin{equation}\label{eq:NS-incomp}
    \nabla \cdot \boldsymbol{u} = 0
\end{equation}
where $\boldsymbol{\sigma}$ is the Cauchy stress tensor and includes both the even and odd stress tensors i.e. $\boldsymbol{\sigma}=\boldsymbol{\sigma^e}+\boldsymbol{\sigma^o}$. The boundary condition on the liquid-gas interface (free surface) is

\begin{equation} \label{eq:bc_free}
    \boldsymbol{\sigma} \cdot \boldsymbol{n_f} = \gamma (\nabla_s \cdot \boldsymbol{n_f}) \boldsymbol{n_f} + \frac{\partial \gamma}{\partial s} \boldsymbol{t_f}
\end{equation}
where $\boldsymbol{n_f}$ and $\boldsymbol{t_f}$ are normal and tangent vectors to the free surface respectively. The boundary conditions on the solid-liquid interface are:
{
\begin{align}
    \boldsymbol{u}\cdot \boldsymbol{n_s} &= 0 \nonumber\\ 
    \boldsymbol{\sigma \cdot n_s} &= -\frac{1}{\beta} \begin{pmatrix} \eta_e & -\eta_o \\ \eta_o & \eta_e \end{pmatrix} (\boldsymbol{u} \cdot \boldsymbol{t_s}) \boldsymbol{t_s}  \label{eq:bc_solid} 
\end{align}
where $\beta$ is the slip length} and $\boldsymbol{n_s}$ and $\boldsymbol{t_s}$ are the normal and tangent vectors to the solid surface. The above equations are no penetration and the Navier-slip conditions respectively with the full viscosity tensor. However, we can show that the odd viscous part of the Navier-slip condition gets evaluated to zero due to the no-penetration condition (see section \ref{sec:weak_formulation}). 

\subsection{The weak formulation} \label{sec:weak_formulation}
To derive the weak form of the Navier-Stokes equations, we choose suitable test functions $\boldsymbol{\nu}$ and $q$, multiply them with equations \ref{eq:NS-eom} and \ref{eq:NS-incomp} respectively and integrate over the computational domain $\Omega$: 

\begin{equation} \label{eq:wf_1}
    \int_\Omega \rho \left(\partial_t \boldsymbol{u} + (\boldsymbol{u}\cdot\nabla)\boldsymbol{u} \right)\cdot \boldsymbol{\nu} d\Omega = \int_\Omega (\nabla \cdot \boldsymbol{\sigma}) \cdot \boldsymbol{\nu} d\Omega + \int_\Omega \boldsymbol{f} \cdot \boldsymbol{\nu} d\Omega
\end{equation}
\begin{equation}\label{eq:wf_2}
    \int_\Omega (\nabla \cdot \boldsymbol{u}) q d\Omega = 0
\end{equation}

\begin{figure}[t] 
\vspace{-5pt}
\begin{center}
\includegraphics[height=3in,width=4in]{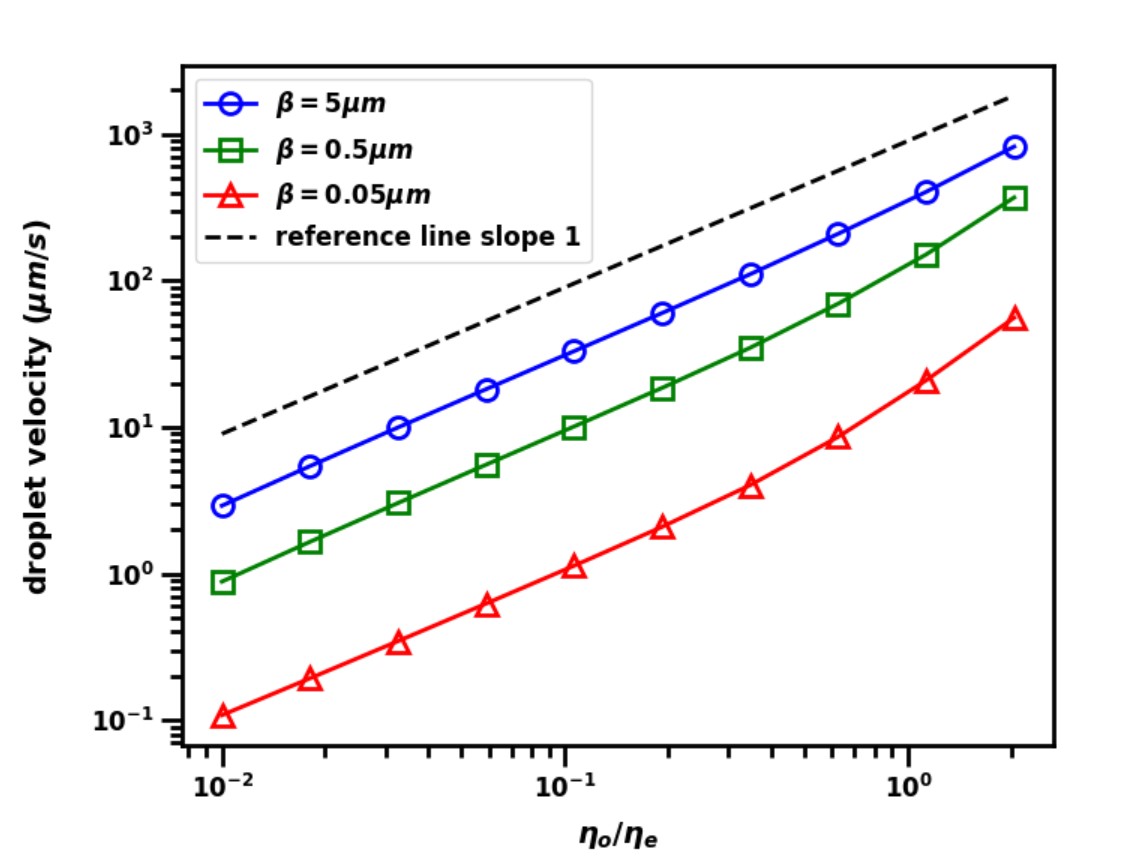}
\end{center}
\vspace{-15pt}
\caption{The effect of slip length on migration velocity
\label{SI_fig1} }
\vspace{-5pt}
\end{figure}

We focus here on the term containing the stress tensor as all other terms have already been implement in COMSOL. For a full derivation of the weak formulation of Navier-Stokes equations, please refer to \cite{ganesan2015,wind2014finite}. From eq. \ref{eq:wf_1}, expanding the term with the stress tensor,

\begin{align}
        \int_\Omega (\nabla \cdot \boldsymbol{\sigma})\cdot \boldsymbol{\nu} &= \int_\Omega \partial_k (\sigma_{ik})\nu_i d\Omega \nonumber\\
        &=  - \int_\Omega \sigma_{ik} \partial_k \nu_i d\Omega +\int_\Omega \partial_k (\sigma_{ik} \nu_i) d\Omega \nonumber\\
        &=  - \int_\Omega (\boldsymbol{\sigma^e}+\boldsymbol{\sigma^o}): (\nabla \boldsymbol{\nu}) d\Omega+ \int_{\partial \Omega}\boldsymbol{\nu} \cdot (\boldsymbol{\sigma^e}+\boldsymbol{\sigma^o} )\cdot \boldsymbol{n} ds \label{eq:weak_form}
\end{align}
where `:' is the is the Frobenious inner product i.e. $\boldsymbol{A}:\boldsymbol{B} = tr(\boldsymbol{A}^T\boldsymbol{B})$ and $\boldsymbol{n}$ is the normal vector to the domain boundary. Analysing the volume integral from eq. \ref{eq:weak_form}, we find that the even stress terms are already implemented by COMSOL. We can therefore isolate the `extra' volumetric term from the weak expression, that is

\begin{align}
    \int_{\Omega} -  \boldsymbol{\sigma^o}: (\nabla \boldsymbol{\nu}) d\Omega &= \int_{\Omega} -  \sigma^o_{ik} \partial_k \nu_i d\Omega\\
    &= \int_{\Omega} - \left(\sigma^o_{xx} \partial_x \nu_1 + \sigma^o_{xy} \partial_y \nu_1 + \sigma^o_{yx} \partial_x \nu_2 + \sigma^o_{yy} \partial_y \nu_2 \right) d\Omega
\end{align}

The test function provided by COMSOL is given by $\nu_i = ( test(u), test(v) )$ and its derivatives $\partial_k \nu_i$ are given by $test(u_x)$, $test(u_y)$, $test(v_x)$ and $test(v_y)$. Therefore, the weak expression is

\begin{equation} \label{eq:extra_weak_contri}
    \int_{\Omega} - \left(\sigma^o_{xx} test(u_x) + \sigma^o_{xy} test(u_y) + \sigma^o_{yx} test(v_x) + \sigma^o_{yy} test(v_y) \right)
\end{equation}
We use COMSOL's \textit{weak contribution} functionality to add this `extra' term to the built-in Navier-Stokes equation. 

Similarly, we analyse the boundary integral from eq. \ref{eq:weak_form} to isolate any `extra' boundary terms that need to be implemented. The boundary integral is split into the free surface boundary $\partial \Omega_f$ and the solid-liquid interface boundary $\partial \Omega_s$, 
\begin{align} \label{eq:boundary_integral}
    \int_{\partial \Omega}\boldsymbol{\nu} \cdot (\boldsymbol{\sigma^e}+\boldsymbol{\sigma^o}) \cdot \boldsymbol{n} ds = \int_{\partial \Omega_f}\boldsymbol{\nu} \cdot (\boldsymbol{\sigma^e}+\boldsymbol{\sigma^o} ) \cdot \boldsymbol{n_f} ds + \int_{\partial \Omega_s}\boldsymbol{\nu} \cdot (\boldsymbol{\sigma^e}+\boldsymbol{\sigma^o} ) \cdot \boldsymbol{n_s} ds
\end{align}
For the free surface integral, substituting boundary conditions from eq. \ref{eq:bc_free}, we get

\begin{equation}
    \int_{\partial \Omega_f} \boldsymbol{\nu} \cdot (\boldsymbol{\sigma^e}+\boldsymbol{\sigma^o} ) \cdot \boldsymbol{n_f} ds = \int_{\partial \Omega_f} \boldsymbol{\nu} \cdot \left(\gamma (\nabla_s . n_f) \boldsymbol{n_f} + \frac{\partial \gamma}{\partial s} \boldsymbol{t_f} \right) ds
\end{equation}
which is the standard boundary condition implemented by COMSOL's laminar flow module on the free surface. On the solid-liquid interface boundary, substituting eq. \ref{eq:bc_solid} in eq. \ref{eq:boundary_integral},  

\begin{align}
    \int_{\partial \Omega_s} \boldsymbol{\nu} \cdot (\boldsymbol{\sigma^e}+\boldsymbol{\sigma^o}) \cdot \boldsymbol{n_s} ds &= \int_{\partial \Omega_s}\boldsymbol{\nu} \cdot \left( -\frac{\eta_e}{\beta} (\boldsymbol{u} \cdot \boldsymbol{t_s}) \boldsymbol{t_s} - \frac{\eta_o}{\beta} \begin{pmatrix} 0 & -1 \\ 1 & 0 \end{pmatrix} (\boldsymbol{u} \cdot \boldsymbol{t_s}) \boldsymbol{t_s} \right)ds \nonumber\\
    &= \int_{\partial \Omega_s}\left[-\frac{\eta_e}{\beta} (\boldsymbol{u} \cdot \boldsymbol{t_s}) (\boldsymbol{\nu} \cdot \boldsymbol{t_s}) - \frac{\eta_o}{\beta}(\boldsymbol{u} \cdot \boldsymbol{t_s}) \left(\boldsymbol{\nu} \cdot \begin{pmatrix} 0 & -1 \\ 1 & 0 \end{pmatrix} \boldsymbol{t_s}\right) \right] ds \label{eq:slip_wf}
\end{align}
Notice that the second term on the right side of the above equation (eq. \ref{eq:slip_wf}) gets evaluated to zero because of the no penetration condition $\boldsymbol{\nu} \cdot \boldsymbol{n_s} = 0$. Therefore, we are left with the following equation

\begin{align}
    \int_{\partial \Omega_s} \boldsymbol{\nu} \cdot (\boldsymbol{\sigma^e}+\boldsymbol{\sigma^o}) \cdot \boldsymbol{n_s} ds = \int_{\partial \Omega_s}\left(-\frac{\eta_e}{\beta} (\boldsymbol{u} \cdot \boldsymbol{t_s}) (\boldsymbol{\nu} \cdot \boldsymbol{t_s}) \right) ds  
\end{align}
which is again the standard Navier-slip condition used by COMSOL's laminar flow module on the solid-liquid interface. Therefore, the weak expression given by eq. \ref{eq:extra_weak_contri} is the only term that needs to be added to the Navier-Stokes equations in COMSOL to model the odd viscous liquid.
In Fig. \ref{SI_fig1} we display the effect of slip length on migration velocity.

\begin{figure}
    \centering
    \includegraphics[width=\linewidth]{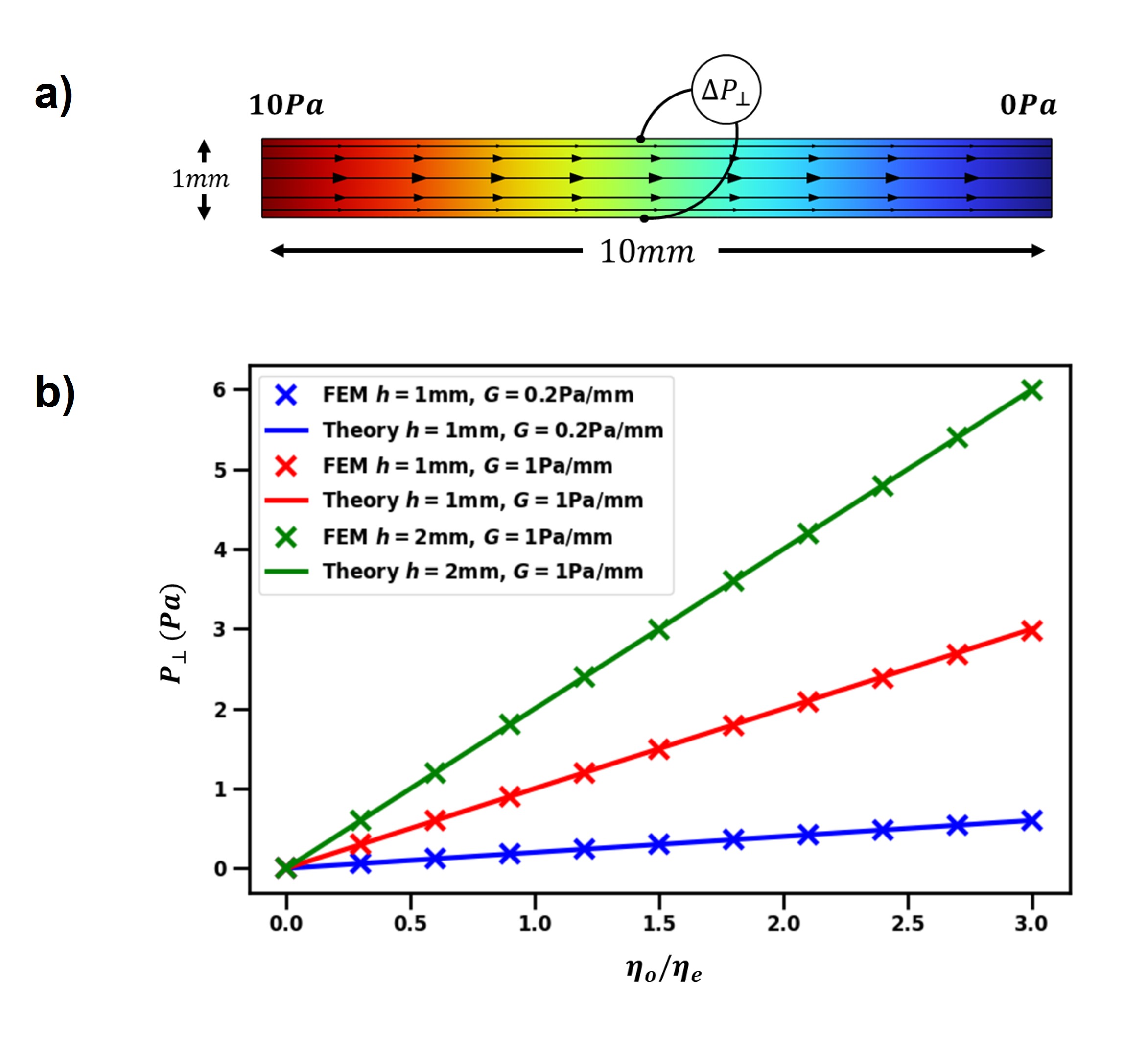}
    \caption{(a) The geometry of a two-dimensional pipe with odd viscous fluid flowing under a constant pressure gradient. (b) Plot of transverse pressure, $P_\perp$ as a function of $\eta_o/\eta_e$ for the Poiseuille flow with odd viscosity. The plot benchmarks the FEM results against the analytical expression for different sets of system parameters. }
    \label{fig:benchmark}
\end{figure}

{The equations of heat transfer are given by:

\begin{equation}
    \rho c_p (\partial_t T + \boldsymbol{u}\cdot \nabla T) = k_{th} \nabla^2 T, 
\end{equation}
where $T$ is the temperature, $c_p$ is the heat capacity, $k_{th}$ is the thermal conductivity of the fluid. The boundary condition used on the fluid-surface interface is $T_0 = T_{gas} + \delta T$, where $\delta T$ was set as $30^\circ C$. On the fluid-gas interface, we use:

\begin{equation}
    k_{th} \boldsymbol{n}\cdot\nabla T = \alpha_{th} (T_\infty - T)
\end{equation}
where $\alpha_{th}$ is the heat transfer coefficient.

The initial conditions used are:
\begin{align}
    \boldsymbol{u} &= 0,  \nonumber\\ 
    p_{init} &= p_{hydrostatic} = \rho g (h-h_{ref}),  \nonumber\\ 
    T_{\infty} &= 293.15K,  \nonumber\\
    T_0 &= T_\infty +\delta T = (293.15 + 30) K = 323.15 K. \nonumber
\end{align}
}

\subsection{Benchmark}
In order to validate the implementation of the model, we benchmark our numerical results against a test problem that can be analytically solved. \cite{vitelli2022} describe the case of two-dimensional Poiseuille flow with odd viscosity through a pipe under a constant pressure gradient. We simulate the system using our model and fig. \ref{fig:benchmark}a shows the geometry of the problem with horizontal pressure gradient, $G = 1 Pa/mm$ and and pipe width, $h = 1 mm$. For a normal fluid ($\eta_o = 0$), the transverse pressure difference across the pipe (labeled as $\Delta P_\perp$ in fig. \ref{fig:benchmark}a) is zero, however, in the presence of odd viscosity ($\eta_o \neq 0$), $\Delta P_\perp$ attains a non-zero value given by \cite{vitelli2022}: 

\begin{equation} \label{eq:benchmark}
    \Delta P_{\perp} = Gh \frac{\eta_o}{\eta_e}
\end{equation} 

Fig. \ref{fig:benchmark}b plots $\Delta P_{\perp}$ as a function of $\eta_o/\eta_e$ for different values of $G$ and $h$. The numerical results from finite element method (FEM) are marked as cross symbols in fig. \ref{fig:benchmark}b  and they match consistently with the analytical solution (eq. \ref{eq:benchmark}) represented as solid lines. This validates the finite element implementation of the odd viscous fluid model.


\def\cprime{$'$}